# "1I/2017 U1 (Oumuamua) Might Be A Cometary Nucleus"

Ignacio Ferrín, Jorge Zuluaga
Solar, Earth and Planetary Physics Group
& Computational Physics and Astrophysics Group (FACom)
Institute of Physics, University of Antioquia
Medellin, Colombia
ignacio.ferrin@udea.edu.co

With the detection of an extra-solar object discovered by the Panstarrs telescope on October 18$^{th}$ 2017, the astrophysics of the minor bodies of our own solar system has jumped to other stars. At the time of writing 14 papers have appeared in a short period of time (30 days), in the Physics Arxiv.org depository, discussing matters like origin (Schneider, 2017; Portegies et al., 2017), photometry (Jewitt et al., 2017; Knight et al., 2017; Bannister et al., 2017; Bolin et al., 2017), spectroscopy (Ye et al., 2017; Masiero, 2017), planetary system formation (Trilling et al., 2017), dark matter (Cyncynates et al., 2017), sending an spacecraft (Hein et al., 2017), consequences of detection (Laughlin and Batygin, 2017), dynamics and kinematics(C. de la Fuente and R. de la Fuente, 2017; Mamajek, 2017), and a detailed model for assessing the origin of interstellar bodies (Zuluaga et al., 2017).

Since the photometry has not been able to detect a coma or a tail, the current consensus is that we are in the presence of an asteroid, whose colors are comparable to those of excited objects of the Kuiper belt or less-red Jupiter Trojans (Bannister et al., 2017), consistent with Kuiper belt colors (Masiero, 2017), colors overlaping the mean colors of D-type Trojan asteroids and other inner solar system populations, and inconsistent with the ultra-red matter found in the Kuiper belt (Jewitt et al., 2017).

In this work we find evidence that the object is of cometary origin.

Four of the above papers have to do with photometric properties (Jewitt et al., 2017; Bolin et al., 2017; Bannister et al., 2017; Knight et al., 2017). Knight et al. do not give colors. Bannister et al. present their results in their Table 1 and 2, but it was not possible to derive Table 2 from Table 1, some information is missing and thus we were unable to use their datasets for this investigation. Jewitt et al. give the colors in the BVRI system. Bolin et al. give the colors in the Sloan Digital Sky Survey photometric system. Using the following data by Bolin et al. :

$$g - r = 0.41 \pm 0.24 \qquad r - i = 0.23 \pm 0.25$$
$$g = 23.5 \pm 0.1 \qquad r = 23.1 \pm 0.1 \qquad i = 22.9 \pm 0.2$$

we convert from the SDSS photometric system to the BVRI system with the transformation equations by Chonis and Gaskell (2008):

$$B = g + (0.327 \pm 0.047)*(g - r) + (0.216 \pm 0.027)$$
$$V = g - (0.587 \pm 0.022)*(g - r) - (0.011 \pm 0.013)$$
$$R = r - (0.272 \pm 0.092)*(r - i) - (0.159 \pm 0.022)$$
$$I = i - (0.337 \pm 0.191)*(r - i) - (0.370 \pm 0.041)$$

obtaining the following BVRI colors of 1I/2017 U1 :

$$B-V = +0.63 \pm 0.49 \qquad V-R = +0.40 \pm 0.42 \qquad R-I = +0.43 \pm 0.43$$

which may be compared with those derived by Jewitt et al.

$$B-V = +0.75 \pm 0.05 \qquad V-R = +0.45 \pm 0.05$$

The agreement is good except that the errors of Jewitt et al. are much smaller. We have been accumulating colors of cometary nuclei (Ferrín, 2006) and we have data for 21 comets. The above colors can be plotted on the color-color diagrams shown in Figure 1, a) B-V vs V-R and b) R-I vs V-R.

The diagrams shows that colors of cometary nuclei fall inside an irregular ellipsoid, but 70% of them fall on a tilted line that we call *the main sequence of cometary nuclei colors*, MS. Plotting the above three observed colors of 1I/2017 U1 on the diagrams shows that the values lie on the MS. This suggests that 1I/2017 U1 is a cometary nucleus.

The next question is if this is an active or an extinct cometary nucleus.

Deep imaging by Meech (2017) failed to show a coma, but we can not exclude the possibility that this was an active comet because there were no observations at or near perihelion (the object was discovered +39 days past perihelion). Some low level cometary nuclei have very short periods of activity. As an example 107P/Wilson-Harrington was active for only 35±5 days (Ferrín et al., 2017).

One implication of this result is that then we do not know from where the object came. It may have come from the inner planetary region, from a local main belt, from the nearby region of a local Jupiter or from the Oort cloud of the parent star.

**Conclusion.** The facts: **a)** both independent datasets agree, **b)** the three data points lie on the MS, **c)** the data point with the smallest error lies almost at the center of the distribution, and **d)** the same result is found in the B-V vs V-R as in the R-I vs B-V diagrams, point to the conclusion that 1I/2017 U1 might be a cometary nucleus.

## ACKNOWLEDGEMENTS

The FACom group is supported by the project "Estrategia de Sostenibilidad 2015 - 2016", sponsored by the Vicerectoría de Investigación of the Universidad of Antioquia, Medellín, Colombia.

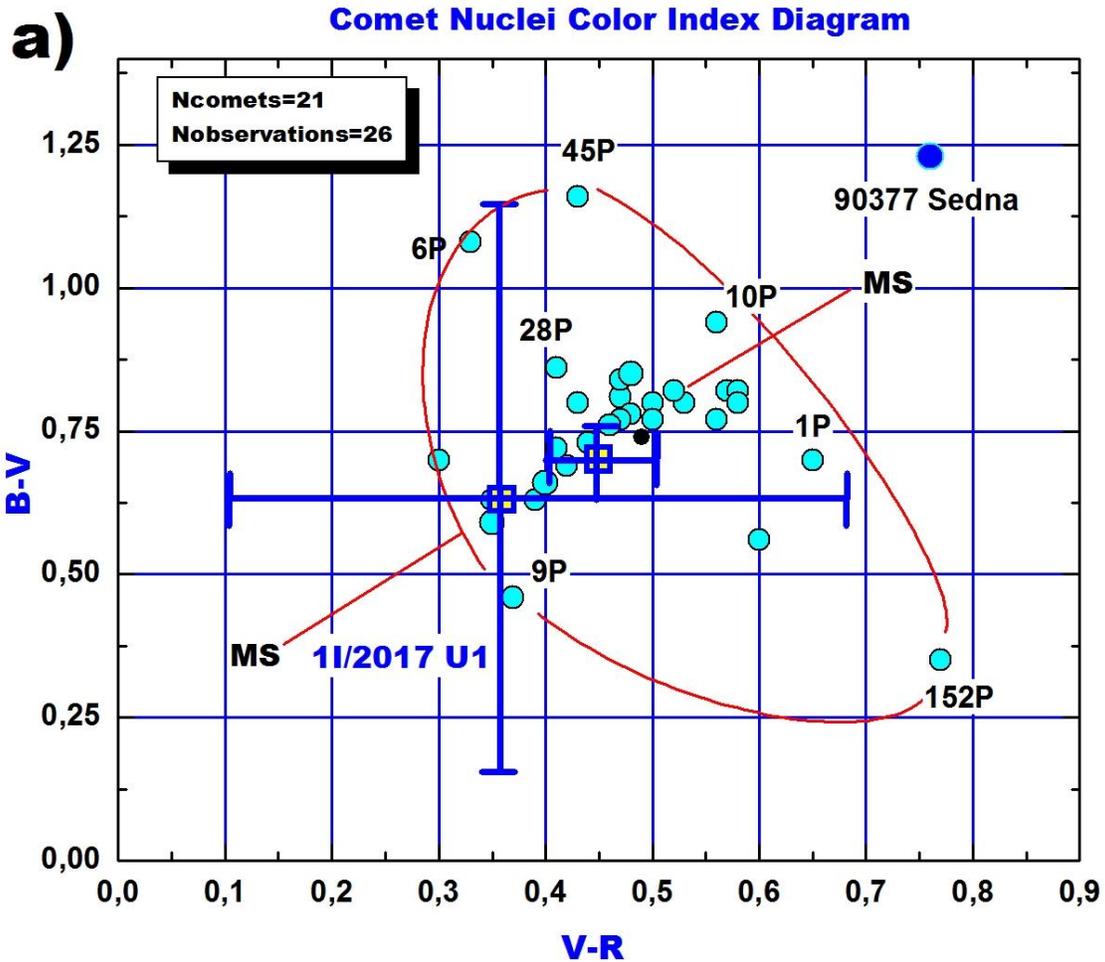

**Figure 1a).** The color-color diagram for cometary nuclei. **a)** B-V vs V-R. The nuclei are located inside an irregular ellipsoid, but 70% of them lie on a tilted line we call *the main sequence of cometary nuclei colors*, MS. Notice that the two measurements of the colors of 1I/2017 U1 by Jewitt et al. (2017) and Bolin et al. (2017), lie on top of the MS. Notice the large error bars of Bolin et al. (2017) data and the small error bars of Jewitt et al. (2017). The black dot represents the centroid of the distribution. The centroid lies inside the error bars of the Jewitt et al. data point.

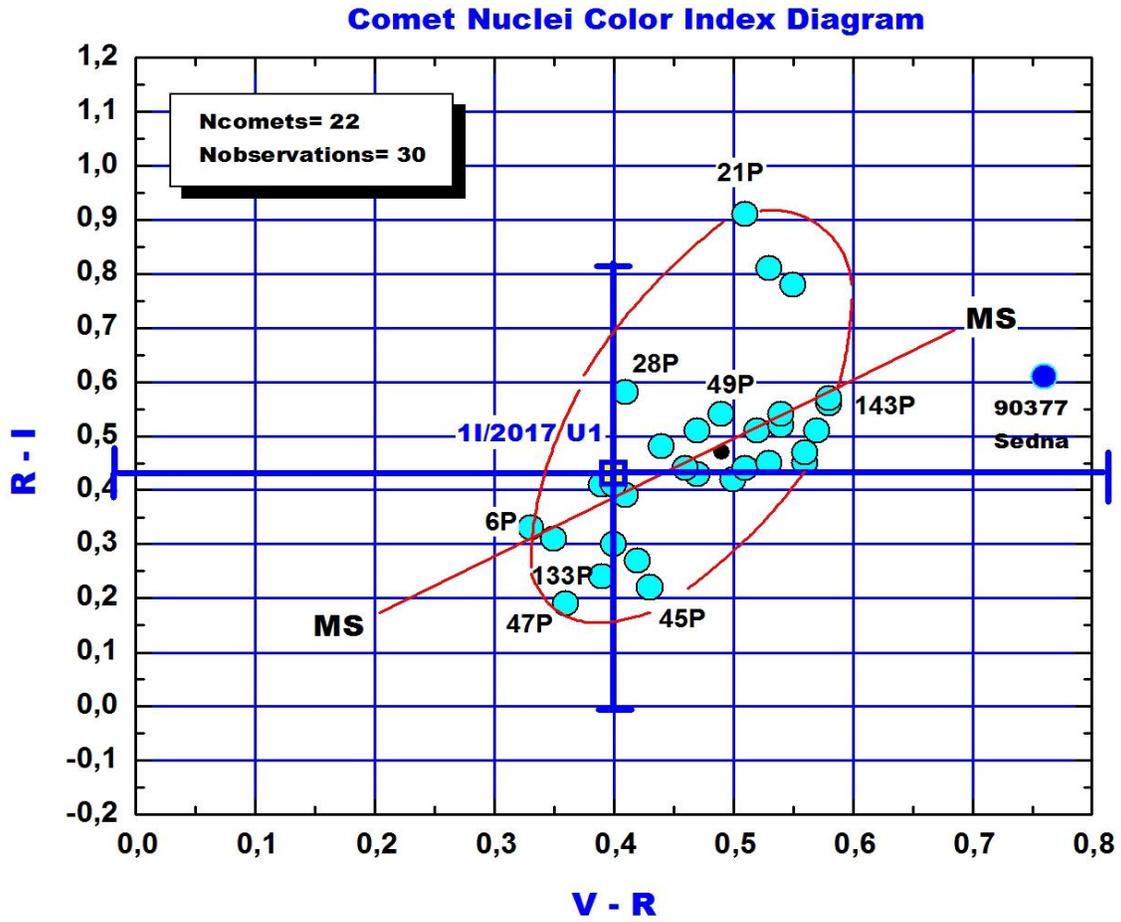

**Figure 1b)** R-I vs V-R. The same comment as in Figure 1a) applies. The black dot represents the centroid of the distribution.